\documentclass[12pt]{article}
\usepackage[margin=1.0in]{geometry}

\DeclareUnicodeCharacter{2212}{-} \DeclareUnicodeCharacter{FB01}{-} \DeclareUnicodeCharacter{FB02}{-}

\usepackage[normalem]{ulem} 

\usepackage{amsmath}
\usepackage{amsfonts}
\usepackage{amssymb}
\usepackage{bm, bbm} 
\usepackage{bbold}
\usepackage{subcaption}
\usepackage{graphicx}
\usepackage{xcolor}
\usepackage{hyperref}
\usepackage{cleveref}
\usepackage{cite}
\usepackage{comment}
\usepackage{multirow}
\usepackage{tikz}
\usetikzlibrary{shapes.geometric, arrows.meta, positioning,decorations.pathreplacing}
\usepackage[T1]{fontenc}
\usepackage[utf8]{inputenc}
\usepackage[greek,english]{babel}

\newcommand{\be}{\begin{equation}}
\newcommand{\ee}{\end{equation}}

\usepackage{xcolor}
\usepackage{bm}
\usepackage{titling}
\usepackage{tikz}
\usetikzlibrary{calc}

\def\bea{\begin{eqnarray}}
\def\eea{\end{eqnarray}}

\title{
Flux Mixing and CP Violation in QCD
}

\author{Motoo Suzuki${}^{1,2,3}$\\[10pt]
{\small\it ${}^1$ SISSA, Via Bonomea 265, Trieste 34136, Italy}\\
{\small\it ${}^2$INFN - Sezione di Trieste, Via Valerio 2, 34127, Italy}\\
{\small\it ${}^3$IFPU, Via Beirut 2, 34014 Trieste, Italy}\\
}

\begin{document}

\maketitle


\begin{abstract}
We argue that kinetic mixing between topological flux sectors generates an effective shift of the QCD $\bar\theta$ angle, thereby inducing CP-violating effects.
To demonstrate this mechanism, we analyze a $(1+1)$-dimensional $U(1)\times U(1)$ gauge theory as a controlled setting, where kinetic mixing leads to observable shifts in electric fluxes. We then extend the analysis to four dimensions using a three-form field description of QCD coupled to an additional $U(1)$ three-form gauge field.
We find that hidden-sector fluxes, through kinetic mixing, shift the effective $\theta$ parameter of QCD and induce a nonzero expectation value of $\langle G\tilde{G}\rangle$.
We discuss the implications for the strong CP problem and clarify under which conditions standard solutions, such as axion or CP/parity-based mechanisms, are compromised or remain robust.

\end{abstract}

\newpage
\tableofcontents
\newpage
\section{Introduction}

The strong CP problem can be viewed as a problem of vacuum selection: why the physical vacuum of QCD preserves CP to a high accuracy. In QCD, the CP/Parity-violating parameter $\bar\theta$ is experimentally constrained to be smaller than $10^{-10}$ from measurements of the neutron electric dipole moment~\cite{Baker:2006ts,Wurm:2019yfj,Abel:2020pzs}. Different values of $\bar\theta$ label distinct superselection sectors of the theory~\cite{tHooft:1976snw,Jackiw:1976pf,Callan:1976je}, which may be viewed as different Universes, as no finite-energy domain wall can connect them. 
Since $\bar\theta$ can take a continuous value in $[0,2\pi)$, the observed Universe corresponds to an extremely small region in this parameter space.

There is no a priori reason to assume that QCD is completely decoupled from additional hidden sectors such as higher-form gauge fields and fluxes. This raises the question of whether such couplings can shift the effective $\theta$-angle and destabilize the CP-conserving vacuum.
To address this question, we consider a minimal extension of QCD in which it is coupled to a three-form $U(1)$ gauge theory through kinetic mixing. In this setup, the local propagating degrees of freedom remain the same as in QCD, while the theory is extended by a sector with nontrivial global structure, characterized by background fluxes and extended objects such as membranes.
The action includes
\begin{align}
S \supset \int \frac{\theta}{8\pi^2}\,{\rm tr}(G\wedge G)
+ \tilde \epsilon\, {\rm tr}(G\wedge G)\,\star dC,
\end{align}
where $C$ is a three-form $U(1)$ gauge field and $\tilde \epsilon$ is a dimensionful kinetic mixing parameter. While this mixing term can preserve CP and parity, a nonzero background flux $\langle \star dC \rangle \neq 0$ induces an effective $\theta$-angle,
\begin{align}
\theta_{\rm eff} \sim \theta + 8\pi^2 \tilde \epsilon \langle \star dC \rangle,
\end{align}
thereby transmitting CP violation from the flux sector into QCD.
However, this conclusion requires some care. Kinetic mixing can in general be removed by field redefinitions, and in the absence of additional couplings it may become unphysical/undetectable. This is well known in four-dimensional $U(1)\times U(1)$ gauge theories~\cite{Holdom:1985ag}, where kinetic mixing between massless gauge fields is unobservable unless they couple to matter fields. It is therefore natural to ask whether this conclusion persists in the presence of topological sectors such as fluxes and three-form gauge fields relevant for QCD. While kinetic mixing is well understood for ordinary gauge fields, its role in topological sectors and its impact on the strong CP problem have not been systematically clarified, to the best of our knowledge.

In this paper, we explore flux mixing and its implications for CP violation in QCD. 
To develop a controlled understanding of this phenomenon, we first study a simpler $(1+1)$-dimensional $U(1)\times U(1)$ gauge theory with kinetic mixing, which serves as an analogue of three-form $U(1)\times U(1)$ gauge theories in four dimensions. 
We identify the conditions under which kinetic mixing becomes physically observable, and clarify the role of charged particles or dual $\eta'$-like fields~\cite{Coleman:1974bu} in rendering the mixing detectable. Quantizing the system, we show that the electric field which is parity-odd is shifted in the presence of a background flux in the other sector.
Based on these results, we investigate flux mixing in a model with a three-form field effective description of QCD. We show that such mixing induces a nonvanishing expectation value of the topological density $\langle G \tilde G \rangle$, thereby contributing to CP-violating observables such as the neutron electric dipole moment.
We further examine the implications for solutions to the strong CP problem. In particular, CP/parity-based solutions~\cite{Nelson:1983zb,Barr:1984qx,Mohapatra:1978fy,Babu:1989rb} do not uniquely select a CP-conserving vacuum and they require additional assumptions about flux mixing.

The organization of this paper is as follows. In Sec.~\ref{sec:review}, we review the $(1+1)$-dimensional toy model and discuss its relation to four-dimensional Yang--Mills theory, setting the stage for the introduction of kinetic mixing. In Sec.~\ref{sec:toy}, we analyze kinetic mixing in $(1+1)$-dimensional $U(1)\times U(1)$ gauge theories and show that the electric flux is shifted by the mixing. In Sec.~\ref{sec:extended_qcd}, we extend this analysis to a three-form model of QCD and discuss its implications for the strong CP problem. 
The final section is devoted to discussion and conclusions.

\section{Preliminaries: $(1+1)d$ Maxwell Theory and Yang--Mills}
\label{sec:review}
\subsection{$(1+1)d$ model}

We start by reviewing $(1+1)$-dimensional QED, which 
provides a simple and calculable toy model capturing features of the low-energy structure of pure Yang--Mills theory at large $N$.
The 2d action is 
\begin{align}
S \supset \int d^2 x \left[
\frac{1}{2g^2} F_{01}^2 + \frac{\theta}{2\pi} F_{01} + A_\mu j^\mu
\right],
\end{align}
where $A_\mu$ is a $U(1)$ gauge field and $F_{01} = \partial_0 A_1 - \partial_1 A_0$ is the field strength. The gauge coupling $g$ has mass dimension one,%
\footnote{The kinetic term is non-renormalizable, so this theory should be regarded as an effective description of some UV completion.}
$j^\mu$ denotes the current of the $U(1)$ charged fields.
Unlike in $(3+1)d$ $U(1)$ gauge theory in the absence of nontrivial topology,
the theta-term in $(1+1)d$ contributes nontrivially to the dynamics discussed below.
The gauge field transforms under the parity $P$, charge conjugation $C$, and time-reversal $T$ as
$P:A_0(x,t) \to A_0(-x,t),~A_1(x,t) \to -A_1(-x,t),~
C:A_\mu(x,t) \to -A_\mu(x,t),~
T:A_0(x,t) \to A_0(x,-t), \quad A_1(x,t) \to -A_1(x,-t)$
which implies
$
P:F_{01}(x,t) \to -F_{01}(-x,t),~
C:F_{01}(x,t) \to -F_{01}(x,t),~
T:F_{01}(x,t) \to F_{01}(x,-t)
$, having transformation properties analogous to those of the electric field in $(3+1)d$. In particular, the $\theta$-term breaks $P$ explicitly.

The theory contains no local propagating degrees of freedom,%
\footnote{In two dimensions, gauge fields carry no local propagating degrees of freedom due to the absence of transverse modes.}
but it admits a global zero mode. 
The dynamics then reduces to quantum mechanics of this zero mode.%
\footnote{See e.g. the lecture note~\cite{tong2018gauge} for more detailed discussion, while we discuss more in later sections after including the kinetic mixing.}
Ignoring matter fields, or taking them to be heavy and integrating them out, the system reduces to pure 2d Maxwell theory.
Upon canonical quantization, the spectrum is labeled by an integer $k \in \mathbb{Z}$, and the energy density eigenvalue,
$
E_k = \frac{1}{2} g^2 \left(\frac{\theta}{2\pi} -  k\right)^2.
$
The corresponding electric field takes the value,
\begin{align}
 F_{01}(k) = g^2\left(k - \frac{\theta}{2\pi}\right).
\end{align}
Physical observables depend only on the combination $k - \theta/2\pi$, which is invariant under the large gauge transformation $\theta \sim \theta + 2\pi$ accompanied by $k \to k+1$.

We refer to the vacua labeled by the integer $k$ as the $k$-vacua~\cite{Witten:1998uka}. 
Different $k$-vacua can be connected by dynamical domain walls, which in two dimensions correspond to charged particles with finite mass. 
In the absence of such charged particles, transitions between different $k$-vacua require infinitely heavy (i.e.\ non-dynamical) domain walls, and hence are forbidden. 
In this case, the distinct $k$-sectors define different Universes following~\cite{Tanizaki:2019rbk} (see also~\cite{Perez-Lona:2025add} for a relation to decomposition~\cite{Hellerman:2006zs,Sharpe:2022ene}).

We can alternatively understand the quantization of the electric field in a different manner.
Under a large gauge transformation, the Wilson line operator transforms as
$
e^{iq \oint_\gamma A} \to e^{iq\oint_\gamma A} e^{i q 2\pi m}=e^{iq\oint_\gamma A},~m\in\mathbb{Z}\ .
$
Requiring gauge invariance of the operator implies the quantization of the electric charge,
$
q
\in \mathbb{Z}\ .
$
On the other hand, we define the dual field strength $F_M$,
$
F_M \equiv \frac{\delta \mathcal{L}}{\delta F_{01}}
= \frac{1}{g^2} F_{01} + \frac{\theta}{2\pi},
$
which is a zero-form scalar, being the dual of the top-form field strength in two dimensions. 
This quantity corresponds to the electric flux (or equivalently, the canonical momentum).
The quantization of electric charge implies that $F_M$ is quantized,
\begin{align}
F_M = q \in \mathbb{Z}.
\end{align}
This reproduces the quantization condition for the electric field,
$
F_{01} = g^2\left(q - \frac{\theta}{2\pi}\right).
$

At $\theta=0$, the Lagrangian is parity invariant. However, in a state with $k \neq 0$, the background electric field, which is odd under the parity, is nonzero,
$
F_{01}(k) = g^2 k\ .
$
Thus, parity is violated in such $k$-sectors due to the non-vanishing electric flux.
The relaxation of the system toward lower-$k$ sectors depends on the presence of domain walls. 
In a constant electric field background, pair production of charged particles (domain walls) can occur via the Schwinger effect~\cite{Brown:1988kg}, allowing transitions $k \to k \pm 1$. If such processes are efficient, the system tends to relax toward the parity-symmetric sector $k=0$. Otherwise, it may remain trapped in a parity-violating state over cosmological timescales.

\subsection{$(3+1)d$ Yang--Mills theory at large $N$}

Let us now consider Yang--Mills theory in the large-$N$ limit of $(3+1)d$. In this limit, the $\theta$-dependence of the vacuum energy takes the form~\cite{Witten:1998uka} 
$
E_k = \frac{\mathcal{X_{\rm YM}}}{8\pi^2} (\theta + 2\pi k)^2 + \mathcal{O}(1/N)
$.
This structure motivates an effective description in terms of a $U(1)$ three-form gauge field $C$ with field strength $F = dC$,
\begin{align}
S = \int \left[
-\frac{1}{2\mathcal{X}_{\rm YM}} F\wedge \star F + \frac{\theta}{2\pi} F
\right]
 + \mathcal{T}\int_M {\rm vol}_3+ q\int_M C,
\end{align}
where 
we have the correspondence $\frac{1}{2\pi} F \sim \frac{1}{8\pi^2} {\rm tr}(G \wedge G)$ with the gluon field strength $G$.
We explicitly include the effective action of a $q$-charged domain wall localized on the worldvolume $M$,%
\footnote{The membrane configurations over which we integrate in the path integral.}
whose tension is $\mathcal{T}$.
The first term for the membrane action 
gives the Nambu--Goto-type worldvolume action, while the second term describes its coupling to the three-form gauge field $C$. The membrane (=domain wall) is dynamical when the tension $\mathcal{T}$ is finite.
The limit of the non-dynamical domain wall corresponds to~$\mathcal{T}^4/\mathcal{\mathcal{X}_{\rm YM}}^3\gg 1$ leading to the exponential suppression of the nucleation rate~
\(
\Gamma  \sim  \exp(-c\, \mathcal{T}^4/\mathcal{X}_{\rm YM}^3)
\)~\cite{Coleman:1977py,Callan:1977pt,Garriga:1993fh}.
This system is analogous to $(1+1)d$ Maxwell theory: the three-form field has no local propagating degrees of freedom while there is the Kogut-Susskind pole~\cite{Kogut:1974kt,Luscher:1978rn}, the $\theta$ parameter is $2\pi$-periodic due to  quantization $\int dC \in 2\pi \mathbb{Z}$, the ``electric field'' is also quantized~\cite{Bousso:2000xa} as in the 2d theory, i.e. $\langle \star dC\rangle = \mathcal{X}_{\rm YM} \left(k - \frac{\theta}{2\pi}\right)\ ,~k \in \mathbb{Z}$ in the limit of non-dynamical domain walls.
The membrane interpolates between vacua with different values of $G\tilde G$. 
Under CP or P, $G\tilde G \to -G\tilde G$, which exchanges the two sides of the wall. 
As a result, a membrane is mapped to a membrane with opposite orientation.

Different $k$-sectors are separated by domain walls/membranes, which are higher-dimensional analogues of charged particles in $(1+1)d$. In the large-$N$ limit, the tension of these domain walls scales with a positive power of $N$~\cite{Witten:1998uka,Shifman:1998if}, and diverges as $N \to \infty$. Consequently, transitions between different $k$-sectors are neglected, and the distinct $k$-vacua behave as different Universes.
At large but finite $N$, however, domain walls have finite tension and can nucleate, allowing transitions between sectors. In this case, the system can dynamically relax toward the lowest energy state.

In QCD, the $k$-vacua was discussed in~\cite{Gabadadze:2002ff} in the heavy quark mass limit, where a three-form effective description of QCD was employed~\cite{DiVecchia:1980yfw,Rosenzweig:1979ay,Nath:1979ik,Nath:1980nf}. For light quark masses and $N_c=3$, the situation is less clear. Nevertheless, in this work we assume that our Universe does not reside in any excited $k$-sector.

\section{Flux mixing in $(1+1)d$ model}
\label{sec:toy}

We consider a $(1+1)$-dimensional Maxwell theory extended by an additional $U(1)$ gauge sector with kinetic mixing, and study how kinetic mixing affects the quantized fluxes. 
The action is
\begin{align}
\label{eq:2d_original_basis}
S = \int d^2 x \Bigg[
&\frac{1}{2g^2} F_{01}^2 + \frac{\theta}{2\pi} F_{01}
+ \frac{1}{2g'^2} {F'_{01}}^2 + \frac{\theta'}{2\pi} F'_{01}
+ \frac{\epsilon}{g g'} F_{01} F'_{01} + A_\mu j^\mu + A'_\mu {j'}^\mu
\Bigg],
\end{align}
where $A_\mu$ and $A'_\mu$ denote two $U(1)$ gauge fields, $F_{01}$ and $F'_{01}$ are their field strengths, $j^\mu$ and ${j'}^\mu$ are the corresponding currents, and $\epsilon$ parametrizes the kinetic mixing. 
We regard the unprimed current $j^\mu$ as defining the visible sector, in the sense that physical observables are probed through this current, while ${j'}^\mu$ defines a hidden sector whose interactions with the visible sector are suppressed for small $\epsilon$. 
The gauge fields $A_\mu$ and $A'_\mu$ are taken in a charge-quantized basis, such that the corresponding electric charges are integer-valued, $q, q' \in \mathbb{Z}$, as seen from the Wilson lines $e^{iq\oint A}$ and $e^{iq'\oint A'}$.

\subsection{Diagonalization of the kinetic terms and detectable kinetic mixing}

We perform field redefinitions consisting of rescalings and rotations of the photon fields, and clarify to what extent the kinetic mixing is physically meaningful. 
We consider two convenient diagonalized bases. Although the Lagrangian takes different forms in each basis, the physical content of the theory remains unchanged.

We first move to a basis where the mixing is absorbed by redefining the visible-sector photon $A_\mu$:
\begin{align}
S &=
\int d^2 x 
\left[ \frac{1}{2g^2}F_{01}^2 + \frac{\theta}{2\pi}F_{01} 
+ \frac{1}{2g'^2} {F'_{01}}^{2} + \frac{\theta'}{2\pi} F'_{01} 
+ \frac{\epsilon}{gg'} F_{01} F'_{01} + A_\mu j^\mu + A'_\mu j'^\mu \right] \\
&= \int d^2 x 
\left[ \frac{1}{2g^2} \left(F_{01} + \frac{\epsilon g}{g'} F'_{01}\right)^2 
+ \frac{1-\epsilon^2}{2g'^2} {F'_{01}}^2
+ \frac{\theta}{2\pi} F_{01} + \frac{\theta'}{2\pi} F'_{01}
+ A_\mu j^\mu + A'_\mu j'^\mu \right] \\
&= \int d^2 x 
\left[ \frac{1}{2} \tilde F_{01}^2 + \frac{1}{2} \tilde {F'_{01}}^2
+ \frac{\theta}{2\pi} g \tilde F_{01} 
+ \frac{g' \theta' - g \epsilon \theta}{2\pi \sqrt{1-\epsilon^2}} \tilde F'_{01}
+ g \tilde A_\mu j^\mu 
+ \frac{1}{\sqrt{1-\epsilon^2}} \tilde A'_\mu \left(g' j'^\mu - g \epsilon j^\mu\right)
\right],
\end{align}
where we define
\begin{align}
\begin{pmatrix}
\tilde A_\mu\\
\tilde A'_\mu
\end{pmatrix}
=
\begin{pmatrix}
\frac{1}{g} & \frac{\epsilon}{g'}\\
0 & \frac{\sqrt{1-\epsilon^2}}{g'}
\end{pmatrix}
\begin{pmatrix}
A_\mu\\
A'_\mu
\end{pmatrix},~
\begin{pmatrix}
A_\mu\\
A'_\mu
\end{pmatrix}
=
\underbrace{
\begin{pmatrix}
g & -\frac{g \epsilon}{\sqrt{1-\epsilon^2}}\\
0 & \frac{g'}{\sqrt{1-\epsilon^2}}
\end{pmatrix}}_{\text{inverse}}
\begin{pmatrix}
\tilde A_\mu\\
\tilde A'_\mu
\end{pmatrix}.
\end{align}

Another basis can be obtained by absorbing the mixing through a redefinition of $A_\mu'$:
\begin{align}
S &=
\int d^2 x 
\left[ \frac{1}{2g^2}F_{01}^2 + \frac{\theta}{2\pi}F_{01}
+ \frac{1}{2g'^2} {F'_{01}}^2 + \frac{\theta'}{2\pi} F'_{01} 
+ \frac{\epsilon}{g g'} F_{01} F'_{01} + A_\mu j^\mu + A'_\mu j'^\mu \right] \\
&= \int d^2 x 
\left[ \frac{1}{2} \tilde F_{01}^2 + \frac{1}{2} \tilde {F'_{01}}^2
+ \frac{\theta g - \theta' \epsilon g'}{2\pi \sqrt{1-\epsilon^2}} \tilde F_{01}
+ \frac{\theta' g'}{2\pi} \tilde F'_{01} 
+ \frac{1}{\sqrt{1-\epsilon^2}} \tilde A_\mu \left(g j^\mu - \epsilon g' j'^\mu \right)
+ g' \tilde A'_\mu j'^\mu \right],
\label{eq:basis_2}
\end{align}
where we define
\begin{align}
\begin{pmatrix}
\tilde A_\mu\\
\tilde A'_\mu
\end{pmatrix}
&=
\begin{pmatrix}
\frac{\sqrt{1-\epsilon^2}}{g} & 0\\
\frac{\epsilon}{g} & \frac{1}{g'}
\end{pmatrix}
\begin{pmatrix}
A_\mu\\
A'_\mu
\end{pmatrix}, \quad
\begin{pmatrix}
A_\mu\\
A'_\mu
\end{pmatrix}
=
\begin{pmatrix}
\frac{g}{\sqrt{1-\epsilon^2}} & 0\\
-\frac{g' \epsilon}{\sqrt{1-\epsilon^2}} & g'
\end{pmatrix}
\begin{pmatrix}
\tilde A_\mu\\
\tilde A'_\mu
\end{pmatrix}.
\end{align}

In the following discussion, we focus on the regime $\epsilon \ll 1$.
The limit $\epsilon \to 1$ is special: the transformation to canonically normalized fields becomes singular, which is reflected in the divergence of factors such as $1/\sqrt{1-\epsilon^2}$.
Physically, this indicates that one linear combination of gauge fields loses its kinetic term and becomes non-dynamical. As a result, the number of propagating degrees of freedom is reduced, and the system effectively describes a single $U(1)$ gauge field.
Although one can formally absorb the factor $1/\sqrt{1-\epsilon^2}$ into a redefinition of $g$ and $g'$, in $1+1$ dimensions these are dimensionful parameters that set the physical scale. Such a redefinition therefore obscures the physical interpretation of the energy scale, rather than simply corresponding to a harmless change of normalization.
From the viewpoint of ultraviolet completion, kinetic mixing is typically generated small as $\epsilon \ll 1$ as we discuss later. For this reason, we restrict ourselves to this regime in the following analysis. The region near $\epsilon = 1$ requires a separate treatment.

We now discuss when the kinetic mixing is detectable (=physical).
When it is physical, the existence of the other sector can, in principle, be inferred from observables.
Let us first consider the case without matter currents, $j^\mu = j'^\mu = 0$. 
In this situation, the notion of ``physical'' versus ``unphysical'' mixing is somewhat ambiguous, since there is no sector that probes the gauge fields. 
Nevertheless, the mixing is regarded as unphysical because the action can be completely diagonalized into two decoupled sectors by a local and invertible field redefinition.
In other words, the mixing parameter is redundant and can be eliminated by a reparametrization of other couplings. For instance, in the basis of Eq.~\eqref{eq:basis_2}, the fields $\tilde F_{01}$ and $\tilde F'_{01}$ are completely decoupled. The mixing parameter appearing in the $\theta$-term of $\tilde F_{01}$ can then be absorbed into a redefinition of the $\theta$ parameter.
Next, we introduce the visible-sector current $j^\mu$. In the same basis, the two sectors remain decoupled, and $j^\mu$ couples only to $\tilde A_\mu$. From the perspective of the visible sector, no additional gauge field is probed, and the mixing remains unobservable.

The situation changes once both sectors are coupled to matter currents. When both $j^\mu$ and $j'^\mu$ are present, there exists no field redefinition that can simultaneously render the gauge fields decoupled from both currents. One can at most diagonalize either the kinetic terms or the current couplings, but not both at the same time. As a consequence, the kinetic mixing cannot be removed from all physical observables and thus becomes physical in the presence of matter in both sectors.

For later convenience, we extend this discussion to the case where the $\theta$ parameters are promoted to dynamical fields.%
\footnote{In 2d, from the duality or bosonization~\cite{Coleman:1974bu}, the charged fermions discussion is directly translated into $\theta$ fields. The discussion here is also applied to 4d case later.}
If only one of the $\theta$ parameters is dynamical field (denoted as $\theta_{\rm dyn}$), and the other sector has neither a current nor a dynamical $\theta'$, the two sectors can still be decoupled, and the mixing remains unphysical.%
\footnote{This can be interpreted as the gauge field $A_\mu$ becoming massive due to the dynamical $\theta$ when $\theta$ is a massless field through the Stuckelberg-like mass term. In this case, one can redefine the remaining gauge field to eliminate the mixing, leading to two decoupled sectors.}
On the other hand, when both $\theta$ and $\theta'$ are dynamical, the two sectors cannot be simultaneously diagonalized, and the mixing becomes physical.
This conclusion holds irrespective of whether the dynamical $\theta$ fields have additional potentials or couplings.

In summary, the kinetic mixing is physical if both sectors are probed, namely when at least one of $\{j^\mu,\, \theta_{\rm dyn}\}$ in the visible sector and at least one of $\{j'^\mu,\, \theta'_{\rm dyn}\}$ in the hidden sector are present.
As long as the kinetic mixing is physical, its effects can be analyzed in any convenient basis.

\subsection{Quantization}
Let us quantize the system.
Taking $A_0=0,~A_0'=0$ gauge and ignore the coupling with matters, the equations of motion give
\begin{align}
& \frac{1}{g^2}\partial_0^2A_1+\frac{\epsilon}{gg'}\partial_0^2A_1'=0,~ \frac{1}{g^2}\partial_0\partial_1A_1+\frac{\epsilon}{gg'}\partial_1\partial_0A_1'=0\ ,\\
  &\frac{1}{g'^2}\partial_0^2A'_1+\frac{\epsilon}{gg'}\partial_0^2A_1=0,~ \frac{1}{g'^2}\partial_0\partial_1A_1'+\frac{\epsilon}{gg'}\partial_1\partial_0A_1=0\ .
\end{align}
Defining new photon fields $A_1^N,~{A'_1}^N$ as
$
  A_1^N=  \frac{1}{g^2}A_1+\frac{\epsilon}{gg'}A_1',~ {A'_1}^N=  \frac{1}{g^2}A_1'+\frac{\epsilon}{gg'}A_1\ ,
$
the EOMs are separated to two parts,
$
    \partial_0^2 A_1^N=0,~ \partial_0\partial_1 A_1^N=0,~
    \partial_0^2 {A_1'}^N=0,~ \partial_0\partial_1 {A_1'}^N=0\ 
$.
The second and fourth equations imply
\(
\partial_1 A_1^N = f(x), ~ \partial_1 {A_1'}^N = f'(x).
\)
However, these correspond to longitudinal modes, which can be removed by the residual gauge symmetry.%
\footnote{After fixing $A_0=0$, the residual gauge transformations satisfy $\partial_0 \alpha(x,t)=0$, and are therefore time-independent.}
Using this freedom, one can set
\(
\partial_1 A_1 = \partial_1 A_1' = 0,
\)
so that only the spatial zero modes remain and $A_1$ and $A_1'$ depend only on time $t$. 
Thus, the theory reduces to quantum mechanics of the zero modes,
\[
\phi = \int_0^{2\pi R} dx\, A_1, \qquad
\phi' = \int_0^{2\pi R} dx\, A'_1,
\]
which are both $2\pi$-periodic.%
\footnote{The spatial manifold is $S^1$ with radius $R$.}
The effective Lagrangian is 
\begin{align}
L = \frac{1}{4\pi g^2 R} \dot\phi^2 + \frac{\theta}{2\pi} \dot\phi
+ \frac{1}{4\pi g'^2 R} \dot\phi'^2 + \frac{\theta'}{2\pi} \dot\phi'
+ \frac{\epsilon}{2\pi R g g'} \dot\phi \dot\phi'.
\end{align}
The canonical momenta are given by
\begin{align}
p \equiv \frac{\partial L}{\partial \dot\phi}
= \frac{1}{2\pi g^2 R} \dot\phi + \frac{\theta}{2\pi} + \frac{\epsilon}{2\pi R g g'} \dot\phi', \quad
p' \equiv \frac{\partial L}{\partial \dot\phi'}
= \frac{1}{2\pi g'^2 R} \dot\phi' + \frac{\theta'}{2\pi} + \frac{\epsilon}{2\pi R g g'} \dot\phi.
\end{align}
Since $\phi$ and $\phi'$ are $2\pi$-periodic variables, their conjugate momenta are quantized, and the Hilbert space is spanned by
\(
\psi_{k,k'} = e^{i k \phi + i k' \phi'}, \qquad k,k' \in \mathbb{Z}.
\)

Now, we obtain the energy density, 
\begin{align}
\mathcal{E} = \frac{1}{8\pi^2 (1 - \epsilon^2)}
\Big[ g^2 (\theta - 2\pi k)^2 + g'^2 (\theta' - 2\pi k')^2
- 2 g g' \epsilon (\theta - 2\pi k)(\theta' - 2\pi k') \Big],
\end{align}
and the background electric fields,
\begin{align}
F_{01}(k,k') &=
\frac{g^2}{1 - \epsilon^2} \Big(k - \frac{\theta}{2\pi}\Big)
- \frac{g g' \epsilon}{1 - \epsilon^2} \Big(k' - \frac{\theta'}{2\pi}\Big), \\
F'_{01}(k,k') &=
- \frac{g g' \epsilon}{1 - \epsilon^2} \Big(k - \frac{\theta}{2\pi}\Big)
+ \frac{g'^2}{1 - \epsilon^2} \Big(k' - \frac{\theta'}{2\pi}\Big).
\end{align}
In a given basis, the presence of kinetic mixing may appear to induce observable contributions to the electric field. However,
this interpretation should be treated with care: by performing redefinitions, the kinetic mixing can be reshuffled into other parameters, such as couplings or $\theta$ angles.
Thus, kinetic mixing itself may be undetectable or degenerate with other parameters in the absence of independent probes. As discussed, the kinetic mixing becomes detectable only when there exist physical probes in both the visible and hidden sectors (see also the next subsection).

The quantization of the canonical momenta can be understood as the quantization of electric fluxes. To see this, consider the Lagrangian
\[
\mathcal{L}=
- \frac{1}{2g^2}F\wedge\star F+\frac{\theta}{2\pi}F
+\frac{\epsilon}{gg'}F\wedge\star F'
-\frac{1}{2g'^2}F'\wedge\star F'+\frac{\theta'}{2\pi}F'\ .
\]
The dual field strengths, defined as the derivatives of the Lagrangian with respect to $F$ and $F'$, are given by
\[
F_M \equiv \frac{\delta\mathcal{L}}{\delta F}
= -\frac{1}{g^2}\star F+\frac{\theta}{2\pi}+\frac{\epsilon}{gg'}\star F', \qquad
F_M' \equiv \frac{\delta\mathcal{L}}{\delta F'}
= -\frac{1}{g'^2}\star F'+\frac{\theta'}{2\pi}+\frac{\epsilon}{gg'}\star F.
\]
In two dimensions, these correspond to electric fluxes, which are quantized,
\[
F_M = k \in \mathbb{Z}, \qquad F_M' = k' \in \mathbb{Z}.
\]
\footnote{See e.g.~\cite{Reece:2023czb}.}
This leads to
\begin{align}
\label{eq:2d_flux_mat}
\begin{pmatrix}
\star F\\
\star F'
\end{pmatrix}
=
\begin{pmatrix}
\frac{g^2}{1-\epsilon^2}\left(\frac{\theta}{2\pi}-k\right)
-\frac{\epsilon gg'}{1-\epsilon^2}\left(\frac{\theta'}{2\pi}-k'\right)\\
-\frac{\epsilon gg'}{1-\epsilon^2}\left(\frac{\theta}{2\pi}-k\right)
+\frac{g'^2}{1-\epsilon^2}\left(\frac{\theta'}{2\pi}-k'\right)
\end{pmatrix}.
\end{align}
In the following, we refer to $k$ and $k'$ as the excitation numbers.

\subsection{Parity violation via flux mixing}
After quantization, the kinetic mixing parameter appears explicitly in the electric fields. 
A nonvanishing electric flux, being odd under parity, signals parity violation. 
In the previous discussion of quantization, we did not introduce any charged matter or dynamical $\theta$ parameters.
Here, we clarify that the parity violation due to the mixing  becomes physically detectable once appropriate probes---namely, charged matter or dynamical $\theta$ fields---are present.

Let us begin with the state labeled by $k = k' = 0$ in the absence of couplings to matter or dynamical $\theta$ parameters. 
In the basis of the redefined primed photon in~\eqref{eq:basis_2}, the electric fields are given by
\begin{align}
\tilde F_{01} &=
\frac{-g \theta + g' \epsilon \theta'}{2\pi\sqrt{1 - \epsilon^2}}\ ,~
\tilde F'_{01}=
 -\frac{g' \theta'}{2\pi(1 - \epsilon^2)} \ .
\end{align}
This shows that the electric fields are sourced by the $\theta$ parameters in this basis. 
The kinetic mixing parameter is degenerate with a reparametrization of the $\theta$ angles, and is therefore redundant or not physically detectable.
Next, let us turn on excited states while still neglecting matter couplings. The electric fields become
\begin{align}
\tilde F_{01} &=
\frac{-g (\theta - 2\pi k) + g' \epsilon (\theta' - 2\pi k')}{2\pi\sqrt{1 - \epsilon^2}}\ ,~
\tilde F'_{01}=
 -\frac{g' (\theta' - 2\pi k')}{2\pi(1 - \epsilon^2)} \ .
\end{align}
The same value of $\tilde F_{01}$ can always be reproduced by an appropriate choice of $(\theta_{\rm mod}, k_{\rm mod})$ satisfying
$
-g (\theta_{\rm mod} - 2\pi k_{\rm mod}) 
= -g (\theta - 2\pi k) + g' \epsilon (\theta' - 2\pi k') \, .
$
Thus, the effect of kinetic mixing remains redundant.
The situation remains unchanged if we introduce charged matter only in the visible sector: although $\tilde F_{01}$ can be relaxed via pair nucleation, the mixing parameter remains undetectable due to the redundancy.%
\footnote{Alternatively, one may introduce charged matter only in the hidden sector. In this case, however, the absence of visible-sector charges removes probes of the visible electric field.}
The situation changes once charged matter is present in both sectors.
Pair nucleation in the hidden sector induces a fractional shift in the electric field detected by the visible sector, thereby providing an observable signature of the hidden sector.
The detection of such an electric field signals parity violation in the Universe.
In summary, the parity violation via the kinetic mixing  is physically meaningful only when both sectors admit dynamical probes as concluded before.

We now assume that the Lagrangian is parity invariant, with $\theta = \theta' = 0$. 
Even in the absence of visible-sector excitations of $k = 0$, a nonzero hidden-sector flux $k' \neq 0$ induces a nonvanishing electric field:
\begin{align}
F_{01}(0,k') = - \frac{g g' \epsilon}{1 - \epsilon^2} \, k' \, .
\end{align}
Since $F_{01}$ is odd under parity, this corresponds to a parity-violating vacuum configuration. 
Thus, although the action respects parity, the vacuum need not: the theory admits parity-violating vacua induced by flux excitations and mixing. 
To avoid such mixing-induced parity violation, one may require either
\begin{itemize}
\item a sufficiently small mixing parameter $\epsilon$, or
\item a dynamical mechanism that relaxes the hidden-sector flux $k'$ to zero.
\end{itemize}
In the following sections, we discuss the kinetic mixing effect in 4d Yang-Mills and QCD,  possible origins of $\epsilon$ in four-dimensional UV completions in Sec.~\ref{sec:UV_completion}, and 
the relaxation of fluxes via axion dynamics in Sec.~\ref{sec:strong_cp}.

\section{Flux mixing in 4d model}
\label{sec:extended_qcd}

 We begin by considering pure $SU(N)$ Yang--Mills theory coupled to a $U(1)$ three-form gauge field $C'_3$ through kinetic mixing. 
 In large N, 
the low energy effective theory is described with
\begin{align}
\label{eq:yang_mixing}
\mathcal{L} = \int
-\frac{1}{2\mathcal{X}} F_4 \wedge \star F_4
+ \frac{\theta}{2\pi} F_4
+ \frac{\epsilon}{\sqrt{\mathcal{X}\mathcal{X}'}}
 F_4 \wedge \star F'_4
- \frac{1}{2\mathcal{X}'} F'_4 \wedge \star F'_4
+ \frac{\theta'}{2\pi} F'_4 \ ,
\end{align}
where $F_4 = dC_3$ and $F'_4 = dC'_3$, the parameters $\mathcal{X}/(4\pi^2)$ and $\mathcal{X}'/(4\pi^2)$ have mass dimension four and relate to the topological susceptibilities.
In addition, we introduce hidden sector membranes.
This theory is structurally identical to the $(1+1)$-dimensional system with kinetic mixing discussed previously. By direct analogy with that analysis, the expectation values of the four-form field strengths are determined by flux quantization and take the form
\begin{align}
\label{eq:electric_4d}
\langle \star F_4(k,k') \rangle &=
\frac{\mathcal{X}}{1 - \epsilon^2}
\Big( \frac{\theta}{2\pi}-k\Big)
- \frac{\sqrt{\mathcal{X}\mathcal{X}'}\,\epsilon}{1 - \epsilon^2}
\Big( \frac{\theta'}{2\pi}-k'\Big), \\
\langle \star F'_4(k,k') \rangle &=
- \frac{\sqrt{\mathcal{X}\mathcal{X}'}\,\epsilon}{1 - \epsilon^2}
\Big(\frac{\theta}{2\pi}-k\Big)
+ \frac{\mathcal{X}'}{1 - \epsilon^2}
\Big(\frac{\theta'}{2\pi}-k'\Big).
\end{align}
The corresponding vacuum energy density is
\begin{align}
\mathcal{E} =
\frac{1}{8\pi^2 (1 - \epsilon^2)}
\Big[
\mathcal{X} (\theta - 2\pi k)^2
+ \mathcal{X}' (\theta' - 2\pi k')^2
- 2 \sqrt{\mathcal{X}\mathcal{X}'}\, \epsilon (\theta - 2\pi k)(\theta' - 2\pi k')
\Big] \ .
\end{align}

We now turn to an effective description of QCD including a three-form field.
Chiral perturbation theory for mesons can be extended by introducing a $U(1)$ three-form gauge field, which provides a description of the large mass of the $\eta'$ meson~\cite{DiVecchia:1980yfw,Rosenzweig:1979ay,Nath:1979ik,Nath:1980nf}.
We consider this effective theory for $N_f=2$, restricting to the $\pi^0$ and $\eta'$ degrees of freedom, which are relevant for our discussion. The effective action is given by
\begin{align}
\label{eq:action_three_form_QCD}
S &\supset \int \frac{1}{2} d\pi^0 \wedge \star d\pi^0
+ \frac{1}{2} d\eta^0 \wedge \star d\eta^0
- \frac{f_\pi^2 B m_u}{2}\left(\frac{\pi^0}{f_\pi}+\frac{\eta^0}{f_\pi}\right)^2
- \frac{f_\pi^2 B m_d}{2}\left(-\frac{\pi^0}{f_\pi}+\frac{\eta^0}{f_\pi}\right)^2
\\
&\quad
+ \frac{2\eta^0}{f_\pi} \frac{F_4}{2\pi}
+ \theta_0 \frac{F_4}{2\pi}
- \frac{1}{2\mathcal{X}_{\rm YM}} F_4 \wedge \star F_4
\ .\nonumber
\end{align}
Here, $\mathcal{X}_{\rm YM}/(4\pi^2)$ denotes the topological susceptibility of pure Yang-Mills theory, $B$ is a low-energy constant related to the quark condensate, with mass dimension one, and $f_\pi\approx 92$\,MeV is the pion decay constant.
The kinetic and mass terms for mesons are expanded.
The heavier mass eigenstate $\eta'$ meson acquires its mass predominantly through its coupling to $F_4$. 
The up and down-quark mass parameter $m_u$ and $m_d$ are real and positive.
If $\eta^0$ does not couple to $F_4$, a massless meson appears in the limit $m_u \to 0$.
However, when the coupling to $F_4$ is present, integrating out the 3-form field $C$ generates a nonzero $\eta'$ mass even in the chiral limit $m_{u,d} \to 0$.
The resulting mass is given by
\(
m_{\eta'}^2
= \frac{4 \mathcal{X}_{\rm YM}}{4\pi^2 f_\pi^2}
= \frac{4}{f_\pi^2}
\left.\frac{\partial^2 E}{\partial \theta_0^2}\right|_{\theta_0=0}
\)
with the vacuum energy density $E$ of pure Yang-Mills,
which reproduces the Witten–Veneziano relation~\cite{Witten:1979vv,Veneziano:1979ec}.

Before including the kinetic mixing, let us check the QCD topological susceptibility $\mathcal{X}_{\rm QCD}$
is reproduced in this system. The equation of motions of the pion, $\eta^0$, and charge quantization condition for the four-form field strength lead to the 4-form flux.
Expanding in the large $\mathcal{X}_{\rm YM}$ 
compared to the scales set by quark masses and pion mass and decay constant parameters,
we obtain at $\theta_0/2\pi\ll 1$,
\begin{align}
   \left\langle \star \frac{F_4}{2\pi}\right\rangle\approx
\theta_0\frac{m_\pi^2f_\pi^2 m_u m_d}{(m_u+m_d)^2}\ ,
   \label{eq:ts_qcd}
\end{align}
where $m_\pi^2=B(m_u+m_d)$ is used. 
This gives the vacuum energy density $E\approx\frac{m_\pi^2f_\pi^2 m_u m_d}{(m_u+m_d)^2}\frac{\theta_0^2}{2}$, reproducing the topological susceptibility of QCD, 
$\partial^2 E/\partial\theta_0^2\approx
\frac{m_\pi^2f_\pi^2 m_u m_d}{(m_u+m_d)^2}
$~\cite{Shifman:1979if}.
One can straightforwardly extend this analysis into the three flavor cases with the strange quark mass $m_s$ and $\eta^8$ meson.

Now, let us introduce the kinetic mixing with an additional flux sector with a three-form $U(1)$ gauge field $C'$. The action is
\begin{align} 
\label{eq:4d_mix_qcd}
S &\supset \int \frac{1}{2} d\pi^0 \wedge \star d\pi^0
+ \frac{1}{2} d\eta^0 \wedge \star d\eta^0
- \frac{f_\pi^2 B m_u}{2}\left(\frac{\pi^0}{f_\pi}+\frac{\eta^0}{f_\pi}\right)^2
- \frac{f_\pi^2 B m_d}{2}\left(-\frac{\pi^0}{f_\pi}+\frac{\eta^0}{f_\pi}\right)^2
\\
&\quad
+ \frac{2\eta^0}{f_\pi} \frac{F_4}{2\pi}
+ \theta_0 \frac{F_4}{2\pi}
- \frac{1}{2\mathcal{X}_{\rm YM}} F_4 \wedge \star F_4
+\frac{\epsilon }{\sqrt{\mathcal{X}_{\rm YM}\mathcal{X}'}}
\frac{1}{2\pi}F_4 \wedge \star F'_4
-\frac{1}{2\mathcal{X}'}F_4'\wedge \star F_4'
+\theta_0'\frac{F_4'}{2\pi}
\nonumber\\
&+ q'\int_M C' + \mathcal{T}'\int_M' {\rm vol}_3,\nonumber
\end{align}
where, in addition to the mixing part, we introduced the hidden sector membrane Nambu-Goto action with $\mathcal{T}'$ tension and the membrane coupling to the three-from field with the charge $q'\in \mathbb{Z}$. 
In the presence of $\eta^0$ in QCD and the membranes in the hidden sector, the kinetic mixing parameter is detectable in principle because both sectors include the probes as in the case of 2d example.
The dual field strength is defined as $\delta \mathcal{L}/\delta F_4,~\delta \mathcal{L}/\delta F_4'$ as before, and the quantization of the electric charge and EOMs lead to the flux,
\begin{align}
\label{eq:flux_4d}
&\left \langle \star 
 \frac{{\rm tr}(G\wedge G)}{8\pi^2} \right\rangle
 \sim\left\langle \star \frac{F_4(k,k')}{2\pi} \right\rangle\nonumber \\
 &\approx
\frac{m_\pi^2f_\pi^2m_um_d }{(m_u+m_d)^2
}
\left(
\theta_0+\epsilon\sqrt{\frac{\mathcal{X}'}{\mathcal{X}_{\rm YM}}}(\theta_0'-2\pi k')
\right).
\end{align}
This is the central result: hidden flux induces visible CP violation.
In the limit $\epsilon\to 0$, this reduces to the QCD result discussed previously. 
Since $G\tilde G$ is a CP-odd operator, a non-vanishing expectation value leads to CP violation in the visible sector.
For instance, this can be probed through physical responses such as $\eta\to \pi^+\pi^-$~\cite{Shifman:1979if,Crewther:1979pi}.
The dynamics of changing $k'$ depends on the UV completion of the hidden sector and may allow for dynamical relaxation.
We also discuss the indication to the neutron electric dipole moment and the strong CP problem in the later section.

\subsection{A UV completion}
\label{sec:UV_completion}

A simple UV origin of the flux mixing arises from integrating out a heavy axion that couples to multiple gauge sectors. 
In addition to Yang-Mills theory or QCD, we consider a hidden $SU(N')$ gauge theory and a heavy axion $\theta_H$ with the action
\begin{align}
S \supset \int 
&(N_H \theta_H + \theta_0)\frac{{\rm tr}(G \wedge G)}{8\pi^2} \,
+ (N_H' \theta_H + \theta'_0)\frac{{\rm tr}(G' \wedge G')}{8\pi^2} + \frac{f^2_H}{2} d\theta_H \wedge \star d\theta_H
- \frac{1}{2} \Lambda_H^4 \, \theta_H\wedge\star\theta_H \, .
\end{align}
Here, $\theta_H$ is assumed to acquire a large mass $m_H^2 \sim \Lambda_H^4/f_H^2$ from additional strong dynamics.
The $\theta_H$ coupling to two sectors can arise, for example, in KSVZ-type constructions with heavy fermions charged under both gauge groups.
When we integrate out $\theta_H$, this generates an effective interaction between the topological densities.
Below the confinement scale of the hidden sector $\Lambda'$, 
using the identification $\frac{1}{8\pi^2} {\rm tr}(G' \wedge G') \;\simeq\; \frac{1}{2\pi} dC_3'$, the effective action becomes
\begin{align}
S \supset \int 
&\theta_0\frac{{\rm tr}(G \wedge G)}{8\pi^2}
- \frac{N_H N_H'}{\Lambda_H^4} 
\, \frac{{\rm tr}(G \wedge G)}{8\pi^2} \, \frac{dC_3'}{2\pi}
- \frac{1}{2\mathcal{X}'} dC_3' \wedge \star dC_3' 
+ \theta'\frac{dC_3'}{2\pi} \, ,
\end{align}
where $\mathcal{X}'$ is the topological susceptibility of the hidden sector, scaling as $\mathcal{X}' \sim \Lambda'^4$.
The mixing parameter is identified as
\begin{align}
\label{eq:mixing_uv}
\epsilon \;\sim\; \frac{N_H N_H'}{(2\pi)^2\Lambda_H^4} \, \sqrt{\mathcal{X}_{\rm YM}\,\mathcal{X}'} \, .
\end{align}
In the hidden sector with large but finite $N_H$, domain walls/membranes with finite tension will arise, connecting vacua labeled by different excitation numbers $k'$. 
For sufficiently large $N_H$, the nucleation rate of these membranes—mediating transitions between different $k'$—is exponentially suppressed in $N_H$~\cite{Witten:1998uka,Shifman:1998if}. 
As a result, membrane nucleation becomes negligible, and each excitation state can be regarded as an almost stable vacuum.

\subsection{Implications for the strong CP problem}
\label{sec:strong_cp}

We examine the implications of flux mixing for the strong CP problem. 
In QCD, the basis-independent CP-violating parameter is given by
\begin{align}
\bar\theta \equiv \theta_0 + \arg \det (M_u M_d),
\end{align}
where $M_u$ and $M_d$ are the up- and down-type quark mass matrices.
In the presence of additional flux sectors, however, the CP violation experienced by QCD is modified through kinetic mixing.  
To capture this effect in a basis-independent manner, we define an effective CP-violating parameter in terms of the vacuum energy density $\mathcal{E}(\bar\theta)$ as
\begin{align}
\bar\theta_{\rm eff}
\equiv
\frac{\partial \mathcal{E}/\partial \bar\theta}{\partial^2 \mathcal{E}/\partial \bar\theta^2}\,.
\end{align}
Here, the numerator corresponds to the topological density,
$\partial \mathcal{E}/\partial \bar\theta = \langle Q_{\rm QCD} \rangle_{\bar\theta}$,
with
$Q_{\rm QCD} \equiv \frac{1}{8\pi^2}\,\star \mathrm{tr}(G \wedge G)$,
while the denominator corresponds to the topological susceptibility,
$\partial^2 \mathcal{E}/\partial \bar\theta^2=\partial \langle Q_{\rm QCD} \rangle_{\bar\theta}/\partial\bar\theta$.
In pure QCD, this definition reproduces $\bar\theta$ in the small-angle regime, since $\langle Q_{\rm QCD} \rangle = \chi_{\rm QCD}\bar\theta + \mathcal{O}(\bar\theta^3)$, where $\chi_{\rm QCD} = \left.\partial^2 \mathcal{E}/\partial \bar\theta^2\right|_{\bar\theta=0}$.
This definition is motivated by the fact that $\langle Q_{\rm QCD} \rangle$ characterizes CP violation in the vacuum.

In the presence of kinetic mixing, $\langle Q_{\rm QCD} \rangle$ receives contributions from the hidden flux sector. 
In the three-form model of Eq.~\eqref{eq:4d_mix_qcd}, we obtain from~\eqref{eq:flux_4d}
\begin{align}
\bar\theta_{\rm eff}
\simeq
\theta_0 
+ \epsilon \sqrt{\frac{\mathcal{X}'}{\mathcal{X}_{\rm YM}}}
(\theta_0' - 2\pi k') \, .
\end{align}
The second term represents CP violation induced from the hidden sector through kinetic mixing. 
When $\mathcal{X}' \gtrsim \mathcal{X}_{\rm YM}$, the mixing parameter $\epsilon$ must be sufficiently small to satisfy the experimental bound from the neutron electric dipole moment, 
$|\bar\theta_{\rm eff}| \lesssim 10^{-10}$~\cite{Baker:2006ts,Wurm:2019yfj,Abel:2020pzs}.
In the UV completion discussed in the previous section, the mixing parameter $\epsilon$ is determined by the hidden strong dynamics as in Eq.~\eqref{eq:mixing_uv}, leading to
\begin{align}
\bar\theta_{\rm eff}
\simeq
\theta_0 
+
N_H N_H'
\frac{\mathcal{X}'}{(2\pi)^2 \Lambda_H^4}
(\theta_0' - 2\pi k') \, .
\end{align}
Thus, the induced CP violation is controlled by the hidden-sector dynamics and the heavy scale $\Lambda_H$, yielding the constraint
\begin{align}
\frac{N_H N_H' \mathcal{X}'}{\Lambda_H^4}
\;\lesssim\;
10^{-10} \, ,
\end{align}
without requiring fine-tuned cancellations.

\subsubsection{CP/Parity solution}
We assume that the theory respects CP (or parity)~\cite{Nelson:1983zb,Barr:1984qx,Mohapatra:1978fy,Babu:1989rb}, such that
\begin{align}
\theta_0 = \theta'_0 = 0,
\end{align}
and that the QCD sector is prepared in the CP-conserving vacuum with $k = 0$.
In the absence of mixing, these conditions ensure $\langle G\tilde G \rangle = 0$ and hence $\bar\theta_{\rm eff}=0$.
In the presence of flux mixing, effective theta parameter is 
\begin{align}
\bar\theta_{\rm eff}
=
2\pi \epsilon\, k'\sqrt{\frac{\mathcal{X}'}{\mathcal{X}_{\rm YM}}}\, .
\end{align}
This shows that CP invariance alone is not sufficient to guarantee the absence of strong CP violation in the presence of flux mixing.
Avoiding the induced CP violation requires additional structure. 
For instance, this can be achieved if the mixing is sufficiently suppressed or if the hidden-sector flux dynamically relaxes to $k'=0$.

\subsubsection{Axion solution}

On the other hand, in the axion solution~\cite{Peccei:1977hh,Weinberg:1977ma,Wilczek:1977pj} where the axion couples to the QCD topological density, the vacuum dynamically adjusts to minimize $\langle G \tilde G \rangle$, analogous to the $\eta'$ solution in the massless quark limit. 
As a result, the cancellation of CP violation is not obstructed by flux mixing.
This can be interpreted as the statement that the effective $\theta$-parameter remains unphysical in the presence of the axion.
However, in the presence of three-form gauge fields or additional strongly coupled sectors, the axion quality problem~\cite{Georgi:1981pu,Kamionkowski:1992mf,Holman:1992us,Barr:1992qq,Ghigna:1992iv,Dine:1992vx,Kallosh:1995hi} may arise due to generic couplings to these sectors. 
Furthermore, the cosmological evolution can be modified if the flux configuration changes through membrane nucleation, which may shift the effective $\theta$-angle during the evolution of the Universe.

\section{Discussion and Conclusions}

An important question is whether a hidden flux sector with membranes can provide a viable solution to the strong CP problem.%
\footnote{
Recent works in~\cite{Kaloper:2025wgn,Kaloper:2025upu} appear to correspond to a special choice of the mixing parameter in the flux-mixing setup
}
To address this, consider pure Yang--Mills theory with kinetic mixing as in~\eqref{eq:yang_mixing}.
For small $\epsilon$, one can choose an appropriate $k'$ such that $\langle \star F_4\rangle \sim \langle G\tilde G\rangle$ is relaxed.
However, this relaxation necessarily induces a nonzero shift in the hidden flux $\langle \star F'_4\rangle$.
As a result, $\langle \star F_4\rangle$ does not vanish at the energy minimum.
This obstruction can be understood in terms of the $(-1)$-form symmetry~\cite{Cordova:2019jnf,Cordova:2019uob,Tanizaki:2019rbk,McNamara:2020uza,Heidenreich:2020pkc}.
In particular, avoiding spontaneous breaking of the $(-1)$-form symmetry implies solving the strong CP problem~\cite{Aloni:2024jpb}.%
\footnote{See~\cite{Garcia-Valdecasas:2024cqn} for a discussion of explicit breaking of the $(-1)$-form symmetry.}
In the present setup, however, the $(-1)$-form symmetry associated with the QCD topological term remains spontaneously broken, and thus the strong CP problem is not resolved.
Instead, the kinetic mixing renders the effective $\theta$ parameter dynamical, allowing it to be scanned in discrete steps via membrane nucleation.
Achieving a sufficiently small CP-violating phase then requires the Universe to reside in a vacuum where the effective $\theta$ is accidentally close to zero.

A complete understanding requires an analysis of membrane cosmology.%
\footnote{A recent work~\cite{Chakraborty:2025lyp} discusses axion cosmology with membranes.}
There are two key differences from ordinary bubble nucleation:
\begin{itemize}
    \item \textbf{Sequential relaxation:} 
    The system may undergo repeated membrane nucleations, gradually reducing the flux rather than transitioning in a single step.

    \item \textbf{Energy dissipation:} 
    Membranes couple primarily to non-propagating three-form fields, and it is less clear how their energy is transferred to radiation.
\end{itemize}
These features point to a rich cosmology, particularly in post-inflationary scenarios, including possible gravitational wave production.
The basic mechanism of membrane nucleation was described by Brown and Teitelboim~\cite{Brown:1988kg} connecting to cosmological constant problem.%
\footnote{
We introduced the hidden sector flux as an initial condition. However, these fluxes may be determined through the membrane nucleation during e.g. inflation.}
We leave a detailed analysis of membrane cosmology and its implications for the strong CP problem to future research.

In the presence of flux mixing, the vacuum selection aspect of the strong CP problem becomes more manifest. A full understanding requires the dynamics of membranes and flux transitions, which remain only partially understood. We hope that the framework developed here provides a useful starting point for further investigations.

\section*{Acknowledgements}
M.S. thanks Eduardo Garc\'{\i}a-Valdecasas for many helpful comments and suggestions at the early stage of this work.
M.S. also thanks Tom Rudelius for drawing attention to the kinetic mixing with three-form gauge fields and its implications for the strong CP problem at the workshop ``The Strong CP Problem and Its Possible Solutions'' in Pollica, Italy.
M.S. would like to acknowledge support by the European Union - NextGenerationEU, in the
framework of the PRIN Project “Charting unexplored avenues in Dark Matter” (20224JR28W). M.S. is supported by the MUR projects 2017L5W2PT.

\bibliographystyle{JHEP}
\bibliography{ref}

\end{document}